\documentstyle[amsmath,epsf,epsfig,preprint,aps,amssymb]{revtex}
\draft \preprint{SNUTP 04-006}
\begin{document}
\title{\Large\bf Trinification with 
$\sin^2\theta_W=\frac38$ and seesaw neutrino mass}
\author{
Jihn E.  Kim\footnote{jekim@phyp.snu.ac.kr}
}
\address{
School of Physics and Center for Theoretical Physics, 
Seoul National University, Seoul 151-747, Korea}
\maketitle

\begin{abstract}
We realize a supersymmetric trinification model with three families
of $({\bf 27}_{\rm tri}+{\bf 27}_{\rm tri}
+\overline{\bf 27}_{\rm tri})$ 
by the $Z_3$ orbifold compactification with two Wilson lines. 
It is possible to break the trinification group to the supersymmetric 
standard model. This model has several interesting features: the
{\it hypercharge quantization}, $\sin^2\theta_W^0=\frac38,$ 
naturally light neutrino masses, and introduction of R-parity.  
The {\it hypercharge quantization} is realized by the choice
of the vacuum, naturally leading toward a supersymmetric 
standard model.  \\
\vskip 0.5cm\noindent [Key words: String orbifold, Trinification, 
$\sin^2\theta_W$, Three families]
\end{abstract}

\pacs{11.30.Hv, 11.25.Mj, 11.30.Ly}

\newpage
\newcommand{\bea}{\begin{eqnarray}}
\newcommand{\eea}{\end{eqnarray}}
\def\beq{\begin{equation}}
\def\eeq{\end{equation}}

\def\one{\bf 1}
\def\two{\bf 2}
\def\five{\bf 5}
\def\ten{\bf 10}
\def\tenb{\overline{\bf 10}}
\def\fiveb{\overline{\bf 5}}
\def\threeb{{\bf\overline{3}}}
\def\three{{\bf 3}}
\def\fb{{\overline{F}\,}}
\def\hb{{\overline{h}}}
\def\Hb{{\overline{H}\,}}

\def\To{${\cal T}_1$\ }
\def\Tt{${\cal T}_2$\ }
\def\Z{${\cal Z}$\ }
\def\ot{\otimes}
\def\tr{{\rm tr}}
\def\sinw{{\sin^2 \theta_W}}

\def\slash#1{#1\!\!\!\!\!\!/}

\newcommand{\dis}[1]{\begin{equation}\begin{split}#1\end{split}\end{equation}}
\newcommand{\beqa}[1]{\begin{eqnarray}#1\end{eqnarray}}

\def\be{\begin{equation}}
\def\ee{\end{equation}}
\def\ben{\begin{enumerate}}
\def\een{\end{enumerate}}
\def\lsl{ l \hspace{-0.45 em}/}
\def\ksl{ k \hspace{-0.45 em}/}
\def\qsl{ q \hspace{-0.45 em}/}
\def\psl{ p \hspace{-0.45 em}/}
\def\ppsl{ p' \hspace{-0.70 em}/}
\def\dsl{ \partial \hspace{-0.45 em}/}
\def\Dsl{ D \hspace{-0.55 em}/}
\def\matrix{ \left(\begin{array} \end{array} \right) }

\def\dag{\dagger}

 \def\NCA{{\em Nuovo Cimento} }
 \def\NIM{{\em Nucl. Instrum. Methods} }
 \def\NIMA{{\em Nucl. Instrum. Methods} A }
 \def\NP{{\em Nucl. Phys.} }
 \def\NPB{{\em Nucl. Phys.} B }
 \def\PL{{\em Phys. Lett.} }
 \def\PLB{{\em Phys. Lett.} B }
 \def\PRL{{\em Phys. Rev. Lett.} }
 \def\PRD{{\em Phys. Rev.} D }
 \def\PR{{\em Phys. Rev.} }
 \def\RMP{{\em Rev. Mod. Phys.} }
 \def\ZPC{{\em Z. Phys.} C }
 \def\PHYSICA{{\em Physica} D }
 \def\CMP{{\em Commun. Math. Phys.} }
 \def\PREP{{\em Phys. Rep.} }
 \def\JMP{{\em J. Math. Phys.} }
 \def\CQG{{\em Class. Quant. Grav.} }
 \def\ANN{{\em Annals of Physics} }
 \def\APP{{\em Acta Phys. Polon.} }
 \def\RPP{{\em Rep. Prog. Phys.} }
 \def\tri{${\bf 27}_{\rm tri}$}

 \def\esix{${\rm E}_6$}
 \def\ee8{${\rm E}_8\times {\rm E}_8^\prime$}
 \def\e8{${\rm E}_8$}

\section{Introduction}

Unification of fundamental forces in the last three
decades \cite{gut} has been a partially successful endeavor.
Probably, the most attractive feature of the unification is the
quantisation of the electromagnetic charge, $Q_{em}({\rm proton})
=-Q_{em}({\rm electron})$. However, the most 
serious problem in this old grand unification(GUT) is the
existence of the fine-tuning problem the so-called gauge hierarchy 
problem. To understand the gauge hierarchy problem, supersymmetry
has been considered, which is then extended
to a consistent superstring theory with a big gauge group
in ten dimensions(10D). One particularly interesting
superstring model is the \ee8\ heterotic
string \cite{ghmr}, because \e8\ contains a chain of
symmetry breaking down to \e8 $\rightarrow$ \esix\ \cite{ex6} and then
down to \esix\ $\rightarrow$ SO(10) $\rightarrow$ SU(5). In this  
\ee8\ heterotic string, the intermediate step \esix\
seems to play a crucial role in the classification of standard 
model(SM) particles. This is because the spinor representation
of SO(10) is included in the fundamental representation
{\bf 27} of \esix. 

For the unification of all fundamental forces, the old GUT idea has 
to be unified with gravity also, which seems to be possible in string 
theory~\cite{gs}. Thus, the unification of all forces is better
studied in the \ee8\ heterotic string.\footnote{
With duality, one can argue that the perturbative heterotic 
string can have other realizations. Here, we stick to the 
perturbative \ee8\ heterotic string.} 
In this theory, there must be a reasonable compactification down to
four dimensions(4D) so that the SM results as an effective theory 
at low energy world. 
The most powerful compactification toward applications in
obtaining 4D effective theories seems to 
be the orbifold compactification \cite{dhvw,inq}. However, the
adjoint representation needed for breaking the GUT group is not
present at the Kac-Moody level $k=1$.\footnote{At higher $k$'s,
it is possible to have an adjoint representation. See, for 
example, \cite{ktye}.} This led to 4D string constructions toward
{\it standard-like} models~\cite{iknq} and flipped SU(5) 
models~\cite{flipped}. 

Here, in obtaining an effective 4D model we include all the
possibilities of assigning the vacuum expectation values(VEV's) 
to Higgs fields. For example, if an SU(5) model does (not) include 
an adjoint representation {\bf 24}, then we say that a SM is (not) 
possible from this model.

However, the standard-like models and the 
flipped SU(5) models suffer from the $\sin^2\theta_W^0(\equiv$ the 
value at the GUT scale) problem toward unification \cite{kim03}. 
The SU(5), SO(10) or \esix\ GUT's with the SM fermions in the 
spinor(or fundamental) representation gives $\sin^2\theta_W^0
=\frac38$, which will be called the ${\rm U(1)_Y}$ hypercharge 
quantization, or simply {\it hypercharge 
quantization}. The $\sin^2\theta_W^0$ problem is 
{the hypercharge quantization} problem.  

The {\it hypercharge quantization problem} can be understood
in the orbifold constructions \cite{kim03} if the 4D gauge group is 
the trinification type, ${\rm SU(3)}^3$ gauge group with 
27 chiral fields(let us define this as ${\bf 27}_{\rm tri}$) in one family, 
suggested in the middle of eighties \cite{glashow}. 
Nevertheless, supersymmetrization of the trinification model does not 
lead to naturally small neutrino masses. Therefore, it was suggested that 
in the supersymmetric trinification one must add another vectorlike 
${\bf 27}_{\rm tri}+ \overline{\bf 27}_{\rm tri}$ \cite{neutrinomass}. 

So far, the trinification with small neutrino masses from the
orbifold compactification was possible
with the bare value of $\sin^2\theta_W^0=\frac14$, where one obtains
just vectorlike lepton humors in addition to
${\bf 27}_{\rm tri}$ \cite{higgsino}, which however does not satisfy the
{ hypercharge quantization}. If the 
{${\rm U(1)_Y}$ hypercharge quantization} is not satisfied, one must
introduce an intermediate scale to fit with data. This is called the
optical unification \cite{giedt}, which depends on details of the 
intermediate scale particles and the magnitudes of the intermediate scales. 
For the  {${\rm U(1)_Y}$ hypercharge quantization}, 
one needs vectorlike
$({\bf 27}_{\rm tri}+ \overline{\bf 27}_{\rm tri})$'s, not just vectorlike
lepton-humor(s) \cite{higgsino}. 

Therefore, for the
{${\rm U(1)_Y}$ hypercharge quantization} it is of utmost importance to
obtain vectorlike
$({\bf 27}_{\rm tri}+ \overline{\bf 27}_{\rm tri})$'s.
In this paper, we fulfil such an objective with an orbifold compactification,
and hence obtain the bare value of $\sin^2\theta_W^0 =\frac38$ naturally.

\section{Trinification with three more (${\bf 27\oplus \overline{27}}$)'s}

Choosing the hypercharge generator as $Y=-\frac12(-2I_1+Y_1+Y_2)$, 
let us denote the trinification spectrum under SU(3)$^3$ as,
\begin{equation}\label{trispectrum}
{\bf 27}_{\rm tri}=
(\threeb,\three,\one)+(\one,\threeb,\three)+(\three,\one,\threeb),
\end{equation}
where
\begin{eqnarray}
{\bf (\bar 3,3,1)}=\Psi_l\longrightarrow
  \Psi_{(\bar M,I,0)}&=& \Psi_{(\bar 1,i,0)}(H_1)_{-\frac12}+
  \Psi_{(\bar 2,i,0)}(H_2)_{+\frac12}
  + \Psi_{(\bar 3,i,0)}(l)_{-\frac12}\nonumber\\
  &+&\Psi_{(\bar 1,3,0)}(N_5)_0+
  \Psi_{(\bar 2,3,0)}(e^+)_{+1}+ \Psi_{(\bar 3,3,0)}(N_{10})_0
\label{rep1}
\\
{\bf (1,\bar 3,3)}=\Psi_q\longrightarrow
  \Psi_{(0,\bar I,\alpha)}\ &=& \Psi_{(0, \bar i,\alpha)}
  (q)_{+\frac16}+ \Psi_{(0,\bar 3,\alpha)}(D)_{-\frac13}
\label{rep2}
\\
{\bf (3,1,\bar 3)}=\Psi_a\longrightarrow
  \Psi_{(M,0,\bar\alpha)}&=& \Psi_{(1,0,\bar\alpha)}
  (d^c)_{\frac13}+ \Psi_{(2,0,\bar\alpha)}(u^c)_{-\frac23}
  + \Psi_{(3,0,\bar\alpha)}(\overline{D})_{+\frac13}
\label{rep3}
\end{eqnarray}
where the representations will be called carrying three 
different {\it humors} as denoted by subscripts: {\it lepton--, quark--,} 
and {\it antiquark--humors}. These names are convenient to 
remember since they contain the designated SM fields. Note that 
{\it lepton--humor} field contains also a pair of Higgs doublets which do 
not carry color charge. With three sets of trinification spectrum, there 
exist three pairs of Higgs doublets.

We take the following orbifold model with two Wilson lines \cite{inq},
\begin{eqnarray}\label{model}
&v =(0~0~0~0~0~\frac13~\frac13~\frac23)(0~0~0~0~0~0~0~0)
\nonumber\\
&a_1 =(0~0~0~~0~0~\frac13~\frac13~\frac23)
(0~0~0~0~0~\frac13~\frac13~\frac23) \\
&a_3 =(\frac13~\frac13~\frac23~0~0~0~0~0)(\frac13~
\frac13~\frac23~0~0~0~0~0)\nonumber
\end{eqnarray}

\medskip

\noindent 
which results to a gauge group $SU(3)^4\times [SU(3)^\prime]^4$.\footnote{
$SU(3)^8$ was considered before in Ref. \cite{su38}} 
The massless fields appear only from the twisted sectors, 
as listed in Table \ref{Table}. 

To obtain all the possible vacuum structure of the compactification,
we consider all the possible VEV's also as commented in the Introduction.
In this spirit, let there exist the following vacuum expectation values 
of the scalar components of the fields appearing in Table \ref{Table},  
\begin{equation}\label{breaking}
\langle {\bf (1,1,1,\three)(1,1,1,1)}\rangle \ne 0,\ \ 
\langle {\bf (1,1,1,1)(1,1,1,\threeb)}\rangle \ne 0,
\end{equation}
so that the last factors in $SU(3)^4$'s, i.e. $SU(3)_4\times SU(3)_{4}
^\prime$, are completely broken, and furhermore, the following link fields 
for identifications of the primed and unprimed $SU(3)$'s,
\begin{eqnarray}
&\langle {\bf (1,\three,1,1)(\three,1,1,1)}\rangle \ne 0,\nonumber\\
&\langle {\bf (1,1,\three,1)(1,\three,1,1)}\rangle \ne 0,\\ \label{link}
&\langle {\bf (\three,1,1,1)(1,1,\three,1)}\rangle \ne 0.\nonumber
\end{eqnarray}
Namely, we identify $\three$ and $\threeb$ of $SU(3)^\prime_1$ as
 $\threeb$ and $\three$ of $SU(3)_2$, 
 $\three$ and $\threeb$ of $SU(3)^\prime_2$ as
 $\threeb$ and $\three$ of $SU(3)_3$, 
 and $\three$ and $\threeb$ of $SU(3)^\prime_3$ as
 $\threeb$ and $\three$ of $SU(3)_1$, respectively. 
Then, the effective theory will be $SU(3)^3$ with the spectrum 
given in Table \ref{tritable}. Note that there results the needed 
three families,
\begin{equation}
3\ [ {\bf 27}_{\rm tri}
 \oplus {\bf 27}_{\rm tri}
 \oplus \overline{{\bf 27}}_{\rm tri}].
\end{equation}

Since the full trinification spectrum is added with the vectorlike 
combination of $ 3\ [ {\bf 27}_{\rm tri}
 \oplus \overline{{\bf 27}}_{\rm tri}]$, 
the bare weak mixing angle is 
$\sin^2\theta_W=\frac38$, fulfilling the hypercharge quantization
\cite{kim03}.

\section{Phenomenology}

There are many indices we deal with:  the untwisted and the twisted sector 
number, the humor(gauge group), and the family indices. So, we use the 
following convention
\begin{equation}\label{indices}
\Psi_{[{\rm family}]}({\rm sector})_{({\rm humor})}.
\end{equation}
For example, $
\Psi_{[{\rm 2}]}({\rm T0})_{({\rm a})}$
represents the second (out of the three) antiquark humor 
$(\three,\one,\threeb)$, 
appearing in the twisted sector T0. This notation will be generalized
to respective fields such as $c^c$, after the symmetry breaking 
$SU(3)^3\rightarrow SU(2)\times U(1)_Y\times SU(3)_c$, and by identifying the 
three remaining light families of fermions. Each twisted sector in
Tables \ref{Table} and \ref{tritable} comes in three copies: so it
is more accurate to represent for example the T0 sectors as T0-1, T0-2, and 
T0-3.
The family indices can be dropped off if unnecessary.

\subsection{Neutrino mass}

The trinification fields of ($ {\bf 27}_{\rm tri} 
\oplus \overline{{\bf 27}}_{\rm tri}$) 
in Table \ref{tritable} can be removed at a large mass scale 
of order $M_G$ by giving VEV's to all singlets in T0
\begin{equation}
\langle \Psi({\rm T0})_{(\one,1,1)}\rangle=M_G.
\end{equation}
The superpotential can be taken as 
\begin{eqnarray}
&g_{ABCDE} \Big[\Psi({\rm T0})_{(\one,1,1)}\Big] \Big[
 \Psi_{[A]}({\rm T2})_{(l)}\Psi_{[B]}({\rm T6})_{(\overline{l})}
 \Psi_{[C]}({\rm T3})_{({\one,1,1})}\Psi_{[D]}({\rm T0})_{({
 \one,1,1})}\Psi_{[E]}({\rm T0})_{({\one,1,1})}\nonumber\\
&+ 
 \Psi_{[A]}({\rm T2})_{(q)}\Psi_{[B]}({\rm T4})_{(\overline{q})}
 \Psi_{[C]}({\rm T5})_{({\one,1,1})}\Psi_{[D]}({\rm T0})_{({
 \one,1,1})}\Psi_{[E]}({\rm T0})_{({\one,1,1})}\nonumber\\
& + \Psi_{[A]}({\rm T2})_{(a)}\Psi_{[B]}({\rm T8})_{(\overline{a})} 
 \Psi_{[C]}({\rm T1})_{({\one,1,1})}\Psi_{[D]}({\rm T1})_{({
 \one,1,1})}\Psi_{[E]}({\rm T3})_{({\one,1,1})}
\Big]
+{\rm h.c.}\label{T2rem}
\end{eqnarray}
where $g_{ABCDE}$ are the couplings and we multiplied three singlet
fields to satisfy the point group selection rule \cite{fiqs}. 

For the case of (\ref{T2rem}),
the three light fermions result from T0. On the other hand,
if we change indices in Eq. (\ref{T2rem}) from $0\leftrightarrow 2$,
then there result light fermions from T2.
Also, a more complicated family structure can be obtained by 
assigning couplings and VEV's judiciously. One can see that (\ref{T2rem})
gives only the Dirac neutrino masses. For a see-saw mechanism, we need
a huge Majorana neutrino mass at high energy scale. So, we consider
the following nonrenormalizable couplings in the superpotential allowed
by $Z_3$ orbifold,
\begin{equation}\label{Maj}
\frac{\lambda_{ABCDEF}}{M^3}\Big[
 \Psi_{[A]}({\rm T0})_{(l)} \Psi_{[B]}({\rm T0})_{(l)}
\Psi_{[C]}({\rm T6})_{(\overline{l})}
 \Psi_{[D]}({\rm T6})_{(\overline{l})}\Psi_{[E]}
({\rm T7})_{(\one,1,1)}\Psi_{[F]}({\rm T7})_{(\one,1,1)}\Big]
\end{equation}
where $M$ is of order the string scale. We will assign huge VEV's to
$\overline{N}_5$'s in T6 and singlets in T7. 
Inserting these VEV's to (\ref{Maj}),
$N_5$'s in T0 obtain huge Majorana masses since the VEV's and $M$
are considered to be of the same order. The $N_5$'s in T0 couple
to light lepton doublets by $N_5({\rm T0})l({\rm T0})H_2({\rm T0})$ 
which render the Dirac neutrino masses of order the electroweak 
scale. Thus, we have all the ingredients needed for the light 
see-saw neutrino masses.

\subsection{R-parity}

The breaking of the trinification gauge group  
down to the standard model gauge group is achieved
by VEV's of $N_{10}$ and $N_5$ directions \cite{higgsino}
(or $\overline{N}_{10}$ and $N_5$, or $\overline{N}_{5}$ and $N_{10}$,
or $\overline{N}_{10}$ and $\overline{N}_5$). Note that 
$\overline{N}_{10}$ and $\overline{N}_5$ appear in the T6 sector. 

Let us choose the VEV's of $N_{10}$ and $\overline{N}_5$, where
$\overline{N}_5$'s appear only in T6. These VEV's certainly break
$SU(3)^3\rightarrow SU(2)_Y\times U(1)\times SU(3)_c$ \cite{kim03}, but
the important thing to note is that 
$\langle\Psi({\rm T6}:\overline{N}_5)_{ (\overline{l})}\rangle$
does not couple $(H_2)_{\frac12}$ field with the lepton doublet 
$(l)_{-\frac12}$, since with the notation given in Eq. (\ref{rep1}),
$\overline{N}_5=\Psi_{(1,\bar 3,0)}, H_2=\Psi_{(\bar 2,i,0)}$, and
$l=\Psi_{(\bar 3,i,0)}$. 
Thus, we obtain a kind of discrete symmetry naturally,
forbidding the mixing of $(H_1)_{-\frac12}$ and $(l)_{-\frac12}$; 
{\it the $R$-parity is introduced} and the proton longevity is 
understood. Certainly, {\it the introduction of this discrete symmetry is by
not allowing VEV's to $N_5$ fields} which would have mixed 
$(H_1)_{-\frac12}$ and $(l)_{ -\frac12}$ if allowed. This is a choice 
of a specific string vacuum from a multitude of vacua. 

The existence of the above discrete symmetry can 
be understood in the following way. The $N_5$ in $(\threeb,\three,\one)$
and $\overline{N}_{5}$ in $(\three,\threeb,\one)$ has the following 
SM quantum numbers in terms of (\ref{rep1}):
$$
N_5: \Psi_{(\bar 1,3,0)} ,\ \ \overline{N}_5: \Psi_{(1,\bar 3,0)}.
$$
Thus, $N_5$ can couple to
\begin{equation}\label{N5coup}
N_5 H_2l,\ \ N_5Dd^c,
\end{equation}
while there is no field which $\overline{N}_5$ can couple to. We assign
a huge VEV to $\langle\overline{N}_5\rangle$, but forbid a VEV of
$N_5$. As stressed before, this is chosen by the string vacuum.
Note, however, there are two sectors(T0 and T6) where $N_5$ appear. With our
example (\ref{T2rem}) the $N_5$'s and $N_{10}$'s in T2 are removed. 
Of course, $N_5$'s in T0 and T2 do not develop VEV's to forbid $H_1-l$ 
mixing. But, all $N_{10}$'s can develop VEV's. Since $N_5({\rm T2})$'s are
removed at a high energy scale, they have
the opposite property from $N_5({\rm T0})$'s.\footnote{Even if more
complicated couplings are introduced, three combinations of 
$N_5$'s remain light.} Thus, the couplings (\ref{N5coup}) can be
considered as the low energy
effective couplings. If we consider the corresponding couplings with 
$N_5({\rm T2})$, they would give
highly suppressed effects at low energy phenomenology. This differentiation 
through the vaccum allows us to assign an effective low energy R-parity 
$R$, to $R(N_5(\rm T0))=-$ and $R(\overline{N}_5({\rm T6}))
=R(N_5({\rm T2}))=+$. On the other hand, $N_{10}$'s in T0 and T2 can
develop huge VEV's, leading to $R(N_{10})=+$. 
Since there is the coupling $N_{10}H_1H_2$, $H_1$ and $H_2$ have the 
same R-parity quantum number. The R-parity of the Higgs fields
must be positive so that their VEV's do not break the R-parity.
From the Yukawa couplings $qu^cH_2$ and
$qd^cH_1$, $u^c$ and $d^c$ must have the same R-parity. From the second 
term of (\ref{N5coup}), $D$ and $d^c$ have the opposite R-parity. 
Also, $H_2$ and $l$ have the opposite R-parity, i.e $l$ has the
negative R-parity, implying $e^c$ also has the negative R-parity. 
Thus, we obtain the
standard R-parity quantum number for leptons,
\begin{equation} 
R(l)=-1, \ \ R(e^c)=-1.
\end{equation}
However, this does not fix the R-parity quantum number of the light
quarks, $q,u^c,d^c$. But it is predicted that the R-parity is opposite
for $d^c$ and $D^c$. 

If $R(D^c)=+1$, then we obtain the standard R-parity quantum numbers,
$R(q,u^c,d^c)=R(l,e^c)=-1$ and proton lifetime from $qqql$ can be made 
sufficiently long.  On the other hand, if $R(D^c)=-1$, then we obtain 
$R(q,u^c,d^c)=-R(l,e^c)=+1$.[Thus, the R-parity can be
considered as the lepton parity.] In this case, $u^cd^cd^{\prime c}$ is
forbidden by the gauge symmetry(not by the R-parity) and $u^cd^cD^{c}$ 
is forbidden by the R-parity. Also, dimension-5 operators in the 
superpotential such as $qqql$ are forbidden by R-parity.\footnote{Only
nonperturbative effects such as by instantons can break this R-parity
at low energy. Then, the dimension-5 operater will be 
sufficiently suppressed.} 
So, it is hopeless to observe proton decay at the current underground
detectors.

\subsection{D-flat directions}

For the low energy N=1 supersymmetry to be valid, there must exist 
F-flat and D-flat directions. It is easy to find F-flat directions. 
For the asymmetric VEV's as we have assigned to $\overline{N}_5$ 
but not to $N_5$, search of D-flat directions is nontrivial. Alas,
we already have so many VEV's for the consistency of our vacuum.
For the D-flatness, we must find at least a direction 
$\Phi^*F_\alpha\Phi=0$ for all $\alpha$, where $\Phi$ is the 
grand column matrix for all the scalar fields and $F_\alpha$ 
are the gauge group generators.  The relevant VEV's for our 
D-flatness are defined as
\begin{eqnarray}
&\langle\Psi({\rm T7})_{(\three,\one,1,1)(\one,1,\three,1)}\rangle
={\rm diag.}(V_{7u}, V_{7d}, V_{7s})\ ,\ \ 
\langle\overline{N}_5\rangle =\langle\Psi_{(1,\bar 3,0)}\rangle
=V_{\overline{N5}}\ ,\nonumber\\
&\langle {N}_{10}\rangle =\langle\Psi_{(\bar 3,3,0)}\rangle=V_{{N10}}\ ,
\ \
\langle\overline{N}_{10}\rangle =\langle\Psi_{(3,\bar 3,0)}\rangle
=V_{\overline{N10}}.\nonumber
\end{eqnarray}
Thus, the conditions for the D-flatness lead to
\begin{eqnarray}
\Phi^\dagger (T_3)_1\Phi&=& \frac12(|V_{\overline{N5}}|^2+|V_{7u}|^2
-|V_{7d}|^2)=0\nonumber\\
\Phi^\dagger(T_3)_2\Phi&=&0\nonumber\\
\Phi^\dagger(Y)_1\Phi&=& \frac13(|V_{\overline{N5}}|^2+2|V_{N10}|^2
-2|V_{\overline{N10}}|^2+|V_{7u}|^2+|V_{7d}|^2-2|V_{7s}|^2)=0\nonumber\\
\Phi^\dagger(Y)_2\Phi&=&\frac23(|V_{\overline{N5}}|^2-|V_{N10}|^2
+|V_{\overline{N10}}|^2)=0\nonumber
\end{eqnarray}
where the subscripts of the generators represent the SU(3) factors 
of the trinification group, and $T_3={\rm diag.}(\frac12,-\frac12,0)$
and $Y={\rm diag.}(\frac13,\frac13,-\frac23)$.
There are enough independent $(|{\rm VEV}|^2)$'s with negative
and positive signs to satisfy the above D-flatness conditions.
 From the above expression, however, our simplified linkage SU(3)$_3^\prime
\rightarrow$SU(3)$^*_1$ in Table \ref{tritable} has more structures such as
$(\one,1,\three,1)^\prime \rightarrow(\threeb,\one,1,1)$, 
but $SU(3)_3^\prime$ and $SU(3)_1$ are completely broken. 

Some string orbifold models contain a mechanism for the doublet-triplet 
splitting by not allowing extra vectorlike quarks but allowing
Higgsinos \cite{iknq}. This has
been reconsidered in field theoretic orbifolds by assigning appropriate
discrete quantum numbers to the bulk fields so that extra massless
vectorlike quarks are forbidden \cite{kawamura}.  
In essence, the string theory interpretation of the doublet-triplet
splitting must arise from the study of the selection rules, summarized
in Ref. \cite{fiqs}. In our case, the doublet-triplet
splitting must occur after the breaking of the trinification gauge group  
down to the standard model gauge group by VEV's of $N_{10},\overline{N}_{10}$,
and $\overline{N}_{5}$. In principle, $\langle{N}_{10}\rangle$ can remove all
the $D-D^c$ fields($\overline{D}-\overline{D}^c$ also) and $H_1-H_2$ fields
($\overline{H}_1-\overline{H}_2$ also). But phenomenologically, we need
just one pair of light Higgsinos of the MSSM, surviving this removal 
process. It is the old $\mu$-problem \cite{kn84} or the MSSM problem 
\cite{higgsino}. At the perturbative level, we have not found such 
a mechanism yet. But, there may be strong dynamics at high energy so 
that the determinant of the Higgsino mass matrix 
vanishes \cite{higgsino}, which we do not pursue here. 
In our case, there is no anomalous U(1) symmetry
from the string compactification since rank 16 is saturated by $SU(3)^8$.
Thus, it is possible to consider the model-independent axion degree
which can translate to a Peccei-Quinn symmetry at low energy \cite{quint}. 
This may help to allow a pair of light Higgs doublets \cite{kn84}.

\section{Conclusion}

We constructed a supersymmetric trinification model with three families
of $({\bf 27_{\rm tri}+27_{\rm tri}+\overline{27}_{\rm tri}})$ by 
the $Z_3$ orbifold compactification with two Wilson lines. It is shown that  
a correct symmetry breaking pattern to the supersymmetric standard model
can be achieved. One of the most attractive features is that the
{\it hypercharge quantization}, i.e. the bare value 
$\sin^2\theta_W^0=\frac38$, is realized by the choice of vacuum. 
It is an important observation
since there is no $Z_3$ orbifold model with any number of
Wilson lines which can directly lead to the
needed spectrum $({\bf 27_{\rm tri}+27_{\rm tri}+\overline{27}_{\rm tri}})$. 
The model presented in this paper gives naturally light neutrino masses, 
and allows an introduction of R-parity. For one choice of the R-charge, 
the D=4 and D=5 baryon number violating operators are excluded, closing the 
window to the proton decay experiment. A natural solution of the 
MSSM problem, however, has to be implemented, which we hope to discuss in
a future communication.

\acknowledgments 
I thank K.-S. Choi, K.-Y. Choi and K. Hwang for useful discussions. 
This work is supported in part by the KOSEF ABRL Grant No. 
R14-2003-012-01001-0, the BK21 program of Ministry of
Education, and Korea Research Foundation Grant No.
KRF-PBRG-2002-070-C00022.

\begin{table}
\begin{center}
\caption{The massless spectrum of the orbifold
(\ref{model}) with the gauge group SU(3)$^8$.}\label{Table}
\begin{tabular}{c|c|c|c}
\hline
sector & twist & multiplicity & massless\ fields \\
\hline
U & &  & None \\
\hline
T0 & $V$ & 9 & $(\one,1,1, 3)(\one,1,1,1)$ \\
  &  & 3 & $(\threeb,\three, \one,1)(\one,1,1,1)+
(\one,\threeb,\three,1)
(\one,1,1,1)+
(\three,\one,\threeb,1)(\one,1,1,1)
$ \\
T1 & $V+a_1$ & 3 & $(\one,1,1,\threeb)(\one,1,1,\three)$ \\
T2 & $V-a_1$ & 9 & $(\one,1,1,1)(\one,1,1,\threeb)$ \\
  &  & 3 & $ (\one,1,1,1)(\threeb,\one,\three,1)+
  (\one,1,1,1)(\three,\threeb,1,1)+
(\one,1,1,1)(1,\three,\threeb,1)
$ \\
T3 & $V+a_3$ & 3 & $(\one,\three,1,1,)(\three,\one,1,1)$ \\
T4 & $V-a_3$ & 3 & $(\one,1,\threeb,1)(\threeb,1,\one,1)$ \\
T5 & $V+a_1+a_3$ & 3 & $(\one,1,\three,1)(\one,\three,1,1)$ \\
T6 & $V+a_1-a_3$ & 3 & $(\one,\threeb,1,1)(\one,1,\threeb,1)$\\
T7 & $V-a_1+a_3$ & 3 & $(\three,\one,1,1)(1,1,\three,1)$\\
T8 & $V-a_1-a_3$ & 3 & $(\threeb,\one,1,1)(\one,\threeb,1,1)$ \\
\hline
\end{tabular}
\end{center}
\end{table}

\begin{table}
\begin{center}
\caption{The massless spectrum with the identification (\ref{link}). 
The gauge group is SU(3)$^3$. The symbol $\{\ \ \}$ in the last column
denotes that some entries are Goldstone bosons and some are heavy ones.
}\label{tritable}
\vskip 0.3cm
\begin{tabular}{c|c|c|c}
\hline
sector & twist & $a_1$  multiplicity & massless\ fields \\
\hline
U &    & None &  \\
\hline
T0 & $V$  & 27 & $\{(\one,1,1)\}$ \\
  &  & 3 & $(\threeb,\three, \one)+
(\one,\threeb,\three)+
(\three,\one,\threeb)$ \\
T1 & $V+a_1$  & 27 & $\{(\one,1,1)\}$  \\
T2 & $V-a_1$  & 27 & $\{(\one,1,1)\}$ \\
  &  & 3 & $
(\threeb,\three,\one)+
  (\one,\threeb,\three)+
(\three,\one,\threeb) $  \\
T3 & $V+a_3$  & 3 & link\ SU(3)$_1^\prime\rightarrow$ SU(3)$_2^*:
 \{\one\oplus {\bf 8}\}$ \\
T4 & $V-a_3$  & 3 & $(\one,\three,\threeb)$  \\
T5 & $V+a_1+a_3$  & 3 & link\ SU(3)$_2^\prime\rightarrow$ 
 SU(3)$_3^*:\{\one\oplus {\bf 8}\}$ 
 \\
T6 & $V+a_1-a_3$ & 3 & $(\three,\threeb,\one)$  \\
T7 & $V-a_1+a_3$ & 3 & link\ SU(3)$_3^\prime
 \rightarrow$ SU(3)$_1^*:\{\one\oplus{\bf 8}\}$ \\
T8 & $V-a_1-a_3$ & 3 & $(\threeb,\one,\three)$  \\
\hline
\end{tabular}
\end{center}
\end{table}

\end{document}